\documentclass[prl,aps,amssymb,twocolumn,showpacs]{revtex4-1}
\usepackage{amsmath}
\usepackage{amssymb}
\usepackage{amsthm}
\usepackage{amsfonts}
\usepackage{listings}
\usepackage{enumerate}
\usepackage{latexsym}

\usepackage{psfrag}

\usepackage{bm}
\usepackage{graphicx}
\usepackage{subfigure}

\newcommand{\beq}{\begin{equation}}
\newcommand{\eneq}{\end{equation}}

\newcommand{\bra}[1]{\left\langle#1\right|}
\newcommand{\ket}[1]{\left|#1\right\rangle}

\input{epsf}

\begin{document}

\tolerance 10000

\newcommand{\vk}{{\bf k}}


\title{Correlation Lengths and Topological Entanglement Entropies of Unitary and Non-Unitary Fractional Quantum Hall Wavefunctions}

\author{B. Estienne$^{1,2}$, N. Regnault$^{3,4}$, and B. A. Bernevig$^3$}
\affiliation{$^1$Sorbonne Universit\'es, UPMC Univ Paris 06, UMR 7589, LPTHE, F-75005, Paris, France}
\affiliation{$^2$CNRS, UMR 7589, LPTHE, F-75005, Paris, France}
\affiliation{$^3$Department of Physics, Princeton University, Princeton, NJ 08544}
\affiliation{$^4$Laboratoire Pierre Aigrain, Ecole Normale Supérieure-PSL Research University, CNRS, Universit\'e Pierre et Marie Curie-Sorbonne Universit\'es, Universit\'e Paris Diderot-Sorbonne Paris Cit\'e, 24 rue Lhomond, 75231 Paris Cedex 05, France}


\begin{abstract}
Using the newly developed Matrix Product State (MPS) formalism for non-abelian Fractional Quantum Hall (FQH) states, we address the question of whether a FQH trial wave function written as a correlation function in a non-unitary Conformal Field Theory (CFT) can describe the bulk of a gapped FQH phase. We show that the non-unitary Gaffnian state  exhibits clear signatures of a pathological behavior. As a benchmark we compute the correlation length of Moore-Read state and find it to be finite in the thermodynamic limit. By contrast, the Gaffnian state has infinite correlation length in (at least) the non-Abelian sector, and is therefore gapless.  We also compute the topological entanglement entropy of several non-abelian states with and without quasiholes. For the first time in FQH the results are in excellent agreement in all topological sectors with the CFT prediction for unitary states. For the non-unitary Gaffnian state in finite size systems, the topological entanglement entropy seems to behave like that of the Composite Fermion Jain state at equal filling.
\end{abstract}

\maketitle

Our understanding of the Fractional Quantum Hall (FQH) effect has benefited substantially from the use of model wavefunctions \cite{Laughlin:1983p301,Halperin83,Jain:1989p294,Moore1991362}. These wavefunctions, although not ground-states of realistic hamiltonians, are nonetheless supposed to capture the universal behavior of the state such as quasiparticle charge, statistics, braiding in the gapped bulk, as well as electron and quasihole exponents on the gapless edge. In a seminal paper\cite{Moore1991362} Moore and Read proposed to use conformal blocks, \emph{i.e.} correlation functions in a Conformal Field Theory (CFT), as a building block to write down bulk model wavefunctions for the ground state and its quasihole excitations. This construction relies on a number of conjectures, the most important being that such a model bulk wavefunction describes a gapped topological state. Another assumption is that the universality class of the fractional quantum Hall state - most notably the braiding and fusion properties of the excitations -  can be read off directly from the bulk CFT. Finally the \emph{bulk-edge correspondence} is usually assumed. It states that (most) properties of the physical gapless edge states should be described by the same CFT that was used to build the bulk wavefunctions. Despite the nontrivial nature of these conjectures, there is a large body of (mostly exact diagonalization) evidence that supports the Moore-Read construction. 

However, this program has been observed recently to break down for non-unitary CFTs. While large sets of bulk trial wavefunctions can be written as correlation functions in a non-unitary CFT\cite{simon2006,Bernevig-PhysRevB.77.184502,Bernevig-PhysRevLett.100.246802,Estienne2010539,Simon-PhysRevB.81.121301}, the bulk and edge CFT can no longer match. Indeed the edge CFT is a low-energy effective theory describing the physical edge states, and as any proper quantum field theory it has to be unitary \cite{Read-PhysRevB.79.245304}. In that case one of the aforementioned hypothesis has to break down : either the edge CFT is different from the one used to write the bulk state - which was shown to be unlikely, at least for the non-unitary Gaffnian state \cite{Read-PhysRevB.79.245304}- or the bulk state itself has to be gapless\cite{Read-PhysRevB.79.045308}. The Gaffnian state is also the prototype of a two dimensional phase where one could study how gapless modes spoil the topological degrees of freedom\cite{Bonderson-PhysRevB.87.195451}.

Unfortunately, a direct numerical observation of the pathology of the non-unitary state as a FQH model wavefunction has been plagued by the relatively small system sizes\cite{simon2006,Regnault-PhysRevLett.101.066803,Toke-PhysRevB.80.205301} that can be reached within exact diagonalization and even with Jack polynomial techniques\cite{Bernevig-2012arXiv1207.3305B}. Thus the question of if and how the non-unitary states fail to be bona fide gapped bulk states has remained unsolved.

Recently, great progress\cite{zaletel-PhysRevB.86.245305,estienne-PhysRevB.87.161112,estienne-2013arXiv1311.2936E} has been made in re-writing many model states using an exact matrix product state\cite{fannes1992finitely,Perez-Garcia:2007:MPS:2011832.2011833} (MPS) description. This description allows for efficient encoding of the states, basically allowing - with excellent accuracy and controlled truncation parameter - the squaring of the sizes previously attained with exact diagonalization. A detailed description of the general method for obtaining the states and their entanglement spectra, approximation parameters, as well as examples of the MPS description of a large number of non-abelian unitary and non-unitary wavefunctions has been provided in Ref.~\onlinecite{estienne-2013arXiv1311.2936E}. 

In this article, we use the MPS machinery to show that model wavefunctions built from non-unitary CFTs exhibit clear signatures of their pathological behavior. We first analyze the topological entanglement entropy \cite{Levin-PhysRevLett.96.110405,Kitaev-PhysRevLett.96.110404} of the unitary Moore-Read and $\mathbb{Z}_3$ Read-Rezayi and show that it accurately matches the CFT prediction both for the ground and for the quasihole states. This is the first time such an agreement is obtained for model FQH states and all topological sectors. Indeed, previous studies have been plagued by finite-size issues\cite{Zozulya-PhysRevB.76.125310} or have not been able to access all the topological sectors\cite{zaletel-PhysRevB.86.245305}. We then perform the same analysis on the non-abelian non-unitary Gaffnian wavefunction, and find that its topological entanglement entropy is that of any abelian state at the same filling $\nu = 2/5$. By adding quasiholes to the Gaffnian wavefunction, we are able to accurately obtain their quantum dimension from topological entropy studies, and find it to be unity. This implies that the quasiholes would be abelian, despite the fact that their fusion rule is non-abelian. This suggests that the bulk of the Gaffnian state is gapless. While the fate of the entanglement entropy in critical one dimensional models based on non-unitary CFTs has been recently discussed\cite{Bianchini-1751-8121-48-4-04FT01}, an analogous study for two dimensional systems was missing until now. We test this by computing the gap of the transfer matrix, which encodes the correlation length of local observables. For the unitary Laughlin, Moore-Read and $\mathbb{Z}_3$ Read-Rezayi state we find a finite correlation length in every abelian and non-abelian sector, while for the non-unitary Gaffnian we find that the correlation length in the quasihole sector extrapolates to infinity in the thermodynamic limit. In the electron sector the extrapolation is equivocal since the Gaffnian correlation length exhibits a level-crossing behavior.

\begin{figure}[htb]
\includegraphics[width=0.92\linewidth]{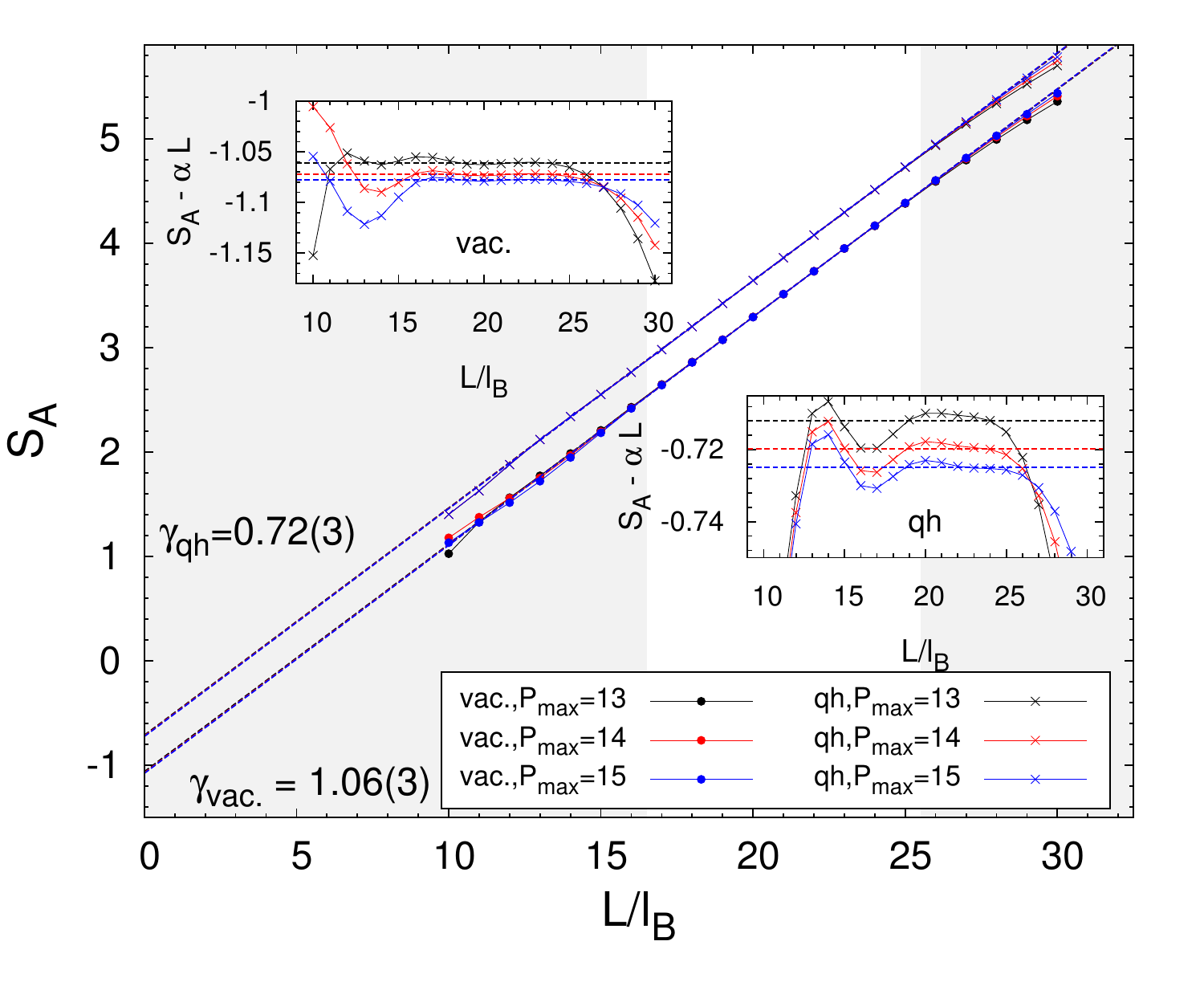}
\caption{Entanglement entropy $S_A$ for the Moore-Read state in the vacuum and sigma sectors (\emph{i.e.} with a non-abelian $\sigma$ quasihole at positions $\pm \infty$) as a function of the cylinder perimeter $L$ and the three largest truncation parameters $P_{\rm max}$ that can be reached. For large $L$, we observe the saturation due to the finite CFT truncation (the saturation increasing with $P_{\rm max}$. $\gamma$ is extracted through a linear fit where we exclude the shaded regions. We consider only perimeters where a convergence of $S_A$ as a function of $P_{\rm max}$ to be lower than $10^{-2}$ has been reached. The extrapolated $\gamma_{vac.}$ (resp. $\gamma_{qh.}$) for the vacuum (resp. sigma) sector is in agreement with the predicted value $\ln {\sqrt{8}}$ (resp. $\ln 2$). {\it Insets}: $S_A$ minus the area law contribution $\alpha L$ for the vacuum (upper left corner) and the sigma (lower right corner) sectors. The value of $\alpha$ which is very close for both topological sectors ($\alpha \simeq 0.22 / l_B$), is extracted from the linear fit.}
\label{ententropypfaffian}
\end{figure}

\begin{figure}[htb]
\includegraphics[width=0.92\linewidth]{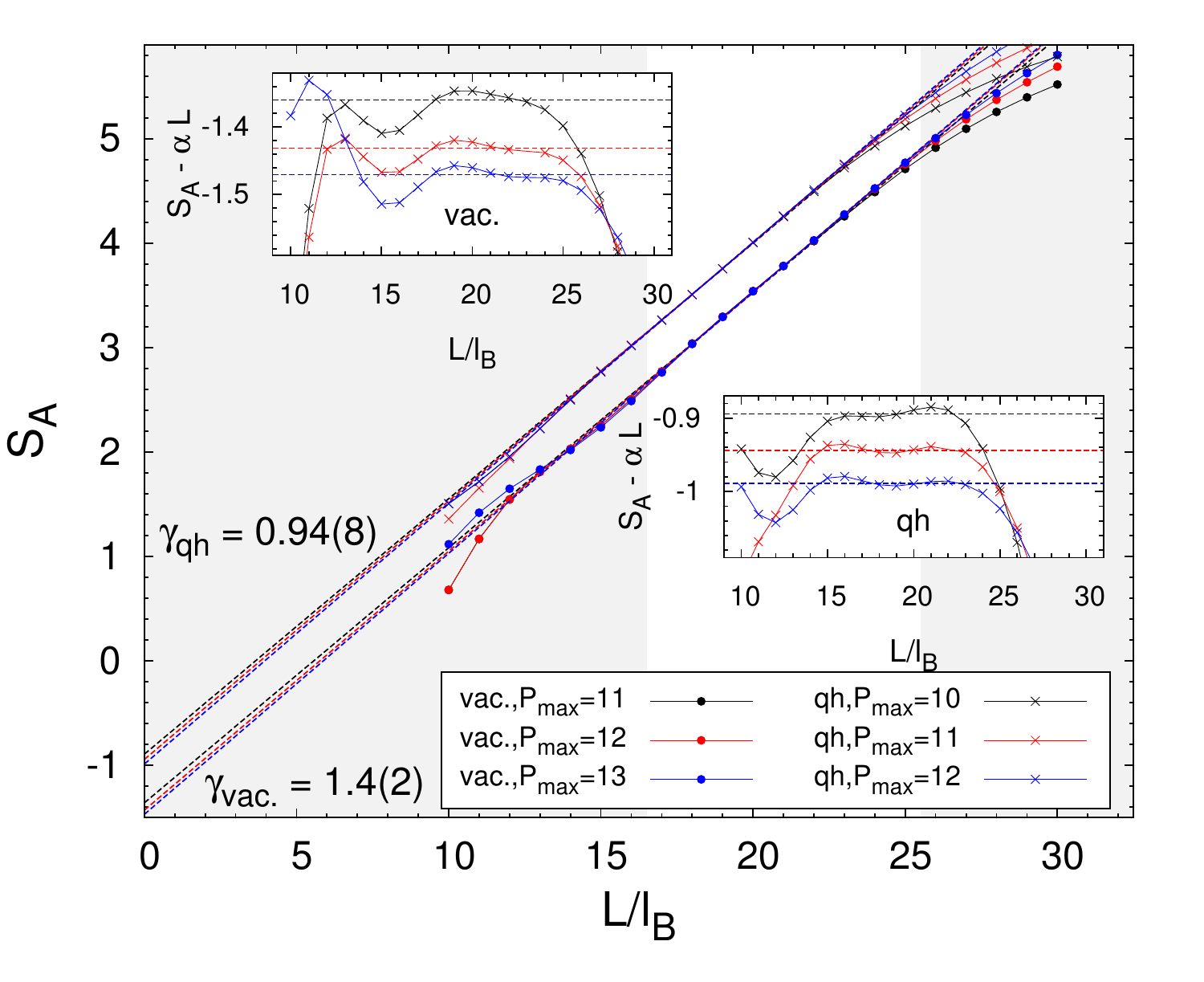}
\caption{Entanglement entropy $S_A$ for the $\mathbb{Z}_3$ Read-Rezayi state in the vacuum sector and with a non-abelian quasihole at each end of the infinite cylinder as a function of the cylinder perimeter $L$ and the three largest truncation parameters $P_{\rm max}$ that can be reached. We use the same conventions than those of Fig.~\ref{ententropypfaffian}, including for the insets.  Note that the lower accuracy on $\gamma$ is due to the data at $P_{\rm max}=11$. Considering only the two largest truncation parameters would give $\gamma = 1.40(3)$. The fitted value of  $\alpha \simeq 0.25 / l_B$ is also very close for both topological sectors.}
\label{ententropyreadrezayi}
\end{figure}

One tool to extract topological information from the ground state wave function is the (topological) entanglement entropy\cite{Levin-PhysRevLett.96.110405,Kitaev-PhysRevLett.96.110404,calabrese-04jsmp06002}. We consider the simplest case of bipartite orbital entanglement on the cylinder between two semi-infinite parts $A$ and $B$ of the system in its ground state $\ket{\Psi}$. This partition is characterized by the reduced density matrix $\rho_A = {\rm Tr}_B \ket{\Psi}\bra{\Psi}$ of subsystem $A$, obtained by tracing out all the $B$ degrees of freedom, a procedure which uses, in the thermodynamic limit, only the highest eigenvalue eigenstate of the MPS transfer matrix, as detailed in Ref.~\onlinecite{estienne-2013arXiv1311.2936E}. Among the various entropies that have been considered as an entanglement measurement, the entanglement entropy is the most popular one. It is defined as the Von Neumann entropy associated with $\rho_A$ \emph{i.e.} $S= -{\rm Tr}_A \left[ \rho_A \ln \rho_A \right]$. For a system in $d$ dimensions with a finite correlation length $\zeta$, the entanglement entropy satisfies the area law\cite{Srednicki-PhysRevLett.71.666}. For two dimensional topological phases, Refs.~\cite{Levin-PhysRevLett.96.110405} and \cite{Kitaev-PhysRevLett.96.110404} showed that the first correction to the area law is a topological term: 
\begin{eqnarray}
S_A&\simeq&\alpha L - \gamma\label{eq:AreaLaw}
\end{eqnarray}
where $L \gg \zeta$ is the length of the boundary of region $A$ (the cylinder perimeter in our case) and $\alpha$ is a non-universal constant. The sub-leading term $\gamma$ is called the topological entanglement entropy: it is a constant for a given topologically ordered phase and is related to the total quantum dimension of the phase. Additional changes in this term appear if regions $A,B$ contain topological (quasihole) excitations, allowing for the determination of the specific quantum dimensions of each topological particle. For a given type of excitations $a$, the quantum dimension $d_a$ defines how the Hilbert space dimension exponentially increases with the number of such excitations. Each type of excitations corresponds to a topological sector. Abelian excitations have a quantum dimension equal to $1$ while non-abelian ones have $d_a > 1$. The topological entanglement entropy for a system with  topological charge $a$ in region A is given by 
\begin{eqnarray}
\gamma &=& \ln \left(\frac{\cal D}{d_a}\right)\label{eq:Gamma}
\end{eqnarray}
where ${\cal D}=\sqrt{\sum_a d_a^2}$ is the total quantum dimension characterizing the topological field theory describing the phase. A state with abelian excitations, at filling factor $\nu=p/q$, such as the Laughlin states or more generally the Jain's composite fermions\cite{Jain:1989p294}, has $q$ abelian topological sectors. Thus its topological entanglement entropy is $\gamma = \log \left(\sqrt{q}\right)$. We numerically obtain the topological entanglement entropy for several states on an infinite cylinder of perimeter $L$. The bi-partition is performed perpendicular to the cylinder axis, such that the length of the boundary between two regions is $L$. Our numerical results contain a truncation parameter, $P_{\text{max}}$, whose meaning is twofold : it is the maximum momentum we use for the edge CFT fields in the auxiliary bond of the MPS \cite{ estienne-2013arXiv1311.2936E}; it is also the maximum descendant field level in the truncated CFT used to build the MPS\footnote{A pictorial representation of the difference between the usual DMRG cutoff and our cutoff, as well as several numerical benchmarks, are presented the Supplementary Material.}.

\begin{figure}[htb]
\includegraphics[width=0.92\linewidth]{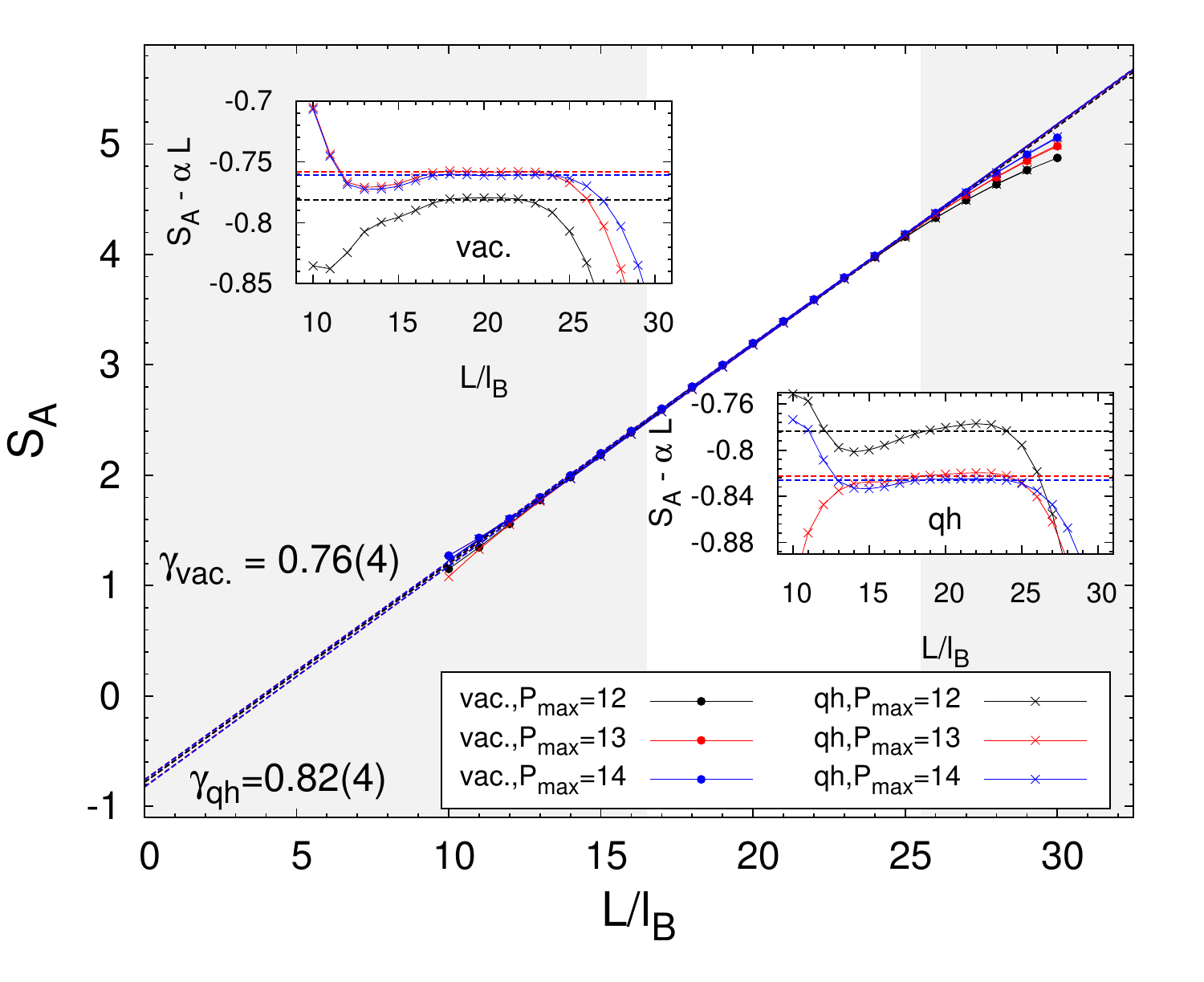}
\caption{Entanglement entropy $S_A$ for the Gaffnian state in the vacuum and sigma sectors as a function of the cylinder perimeter $L$ and the three largest truncation parameters $P_{\rm max}$ that can be reached. We use the same conventions than those of Fig.~\ref{ententropypfaffian}, including for the insets. The fitted value of  $\alpha \simeq 0.20 / l_B$ is also very close for both topological sectors. Note that the difference of the entanglement entropies between the two sectors at any perimeter (in the converged region) is always lower than 0.02.}
\label{ententropyGaffnian}
\end{figure}

\begin{figure}[htb]
\includegraphics[width=0.9\linewidth]{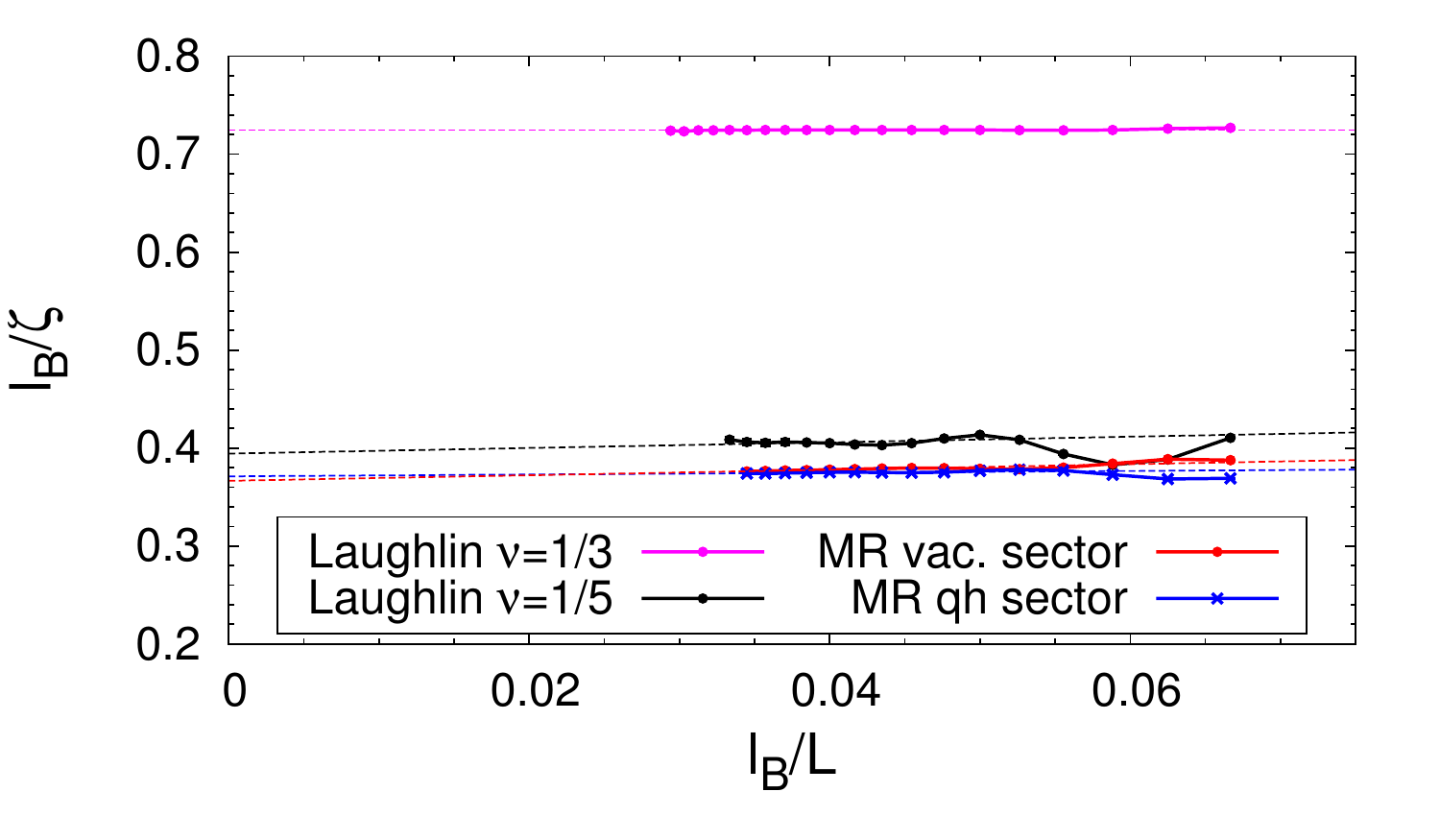}
\caption{Correlation lengths $\zeta$ of the Laughlin $\nu=1/3$, $\nu=1/5$, and the Moore-Read state for both the vacuum and the quasihole sectors as a function of cylinder perimeter $L$. A linear fit as a function of $1/L$ gives the thermodynamical values $\zeta_{3}/l_B=1.381(1)$ for $\nu=1/3$, $\zeta_{5}/l_B=2.53(7)$ for $\nu=1/5$, $\zeta_{\rm vac}/l_B=2.73(1)$ for the Moore-Read vacuum sector and $\zeta_{\rm qh}/l_B=2.69(1)$ for the Moore-Read quasihole sector.}
\label{corrlengthlaughlinmr}
\end{figure}

For the Laughlin state, the topological entanglement entropy has been computed using a MPS representation in Ref.~\onlinecite{zaletel-PhysRevB.86.245305} and matches the theoretical prediction. Beyond abelian states, the simplest example of a non-abelian state is the Moore-Read state\cite{Moore1991362}. It has 6 topological sectors, $4$ abelian ones and $2$ non-abelian ones with a quantum dimension $d_{\sigma}=\sqrt{2}$. In Fig.~\ref{ententropypfaffian}, we show the entanglement entropy as a function of the cylinder perimeter for a for the Moore-Read state in the abelian (or vacuum) sector and the sigma sector. We first focus on the vacuum sector. Two size effects are present in the raw data: for very small perimeters, we are in a regime where the Moore-Read state is not fully developed (the "thin torus" regime). For very large perimeters, the entanglement entropy saturates due to the finite truncation in the CFT. We hence present only the regime where a clear area law of the entanglement entropy is observed. We extrapolate to obtain the topological entanglement entropy $\gamma_{\rm vac.}$ and find it in excellent agreement with the conjectured value $\ln {\sqrt{8}} \simeq 1.039$ for the groundstate. Adding a quasihole at each end of the infinite cylinder to access the sigma (i.e. quasihole) sector and extracting the corresponding $\gamma_{\rm qh}$ from Fig.~\ref{ententropypfaffian} give a non-abelian quasihole quantum dimension of $d_\sigma=1.4$, again in excellent agreement with the CFT conjecture. Note that as expected, the area law linear factor $\alpha$ given by Eq.~\ref{eq:AreaLaw} is identical (within numerical accuracy) in both topological sectors. This result holds true for each state we have considered. Fig.~\ref{ententropyreadrezayi} shows the data for the $\mathbb{Z}_3$ Read-Rezayi state. We again obtain excellent agreement for the topological entanglement entropies with the CFT predictions of $\gamma_{\rm vac.}=1.44768$ and $\gamma_{\rm qh}=0.96647$ for the ground-state and quasihole sectors, respectively.  

In Fig.~\ref{ententropyGaffnian} we present the results for the non-unitary Gaffnian state, both in the ground-state and with a $\sigma$ particle in side $A$. We find a clear area law in both cases. Upon extrapolation, the topological entanglement entropy is equal (within numerical error) between these two cases, suggesting that the quantum dimension of the $\sigma$ particle is $1 \pm 0.02$. This would correspond to an abelian particle although its fusion rules \cite{estienne-2013arXiv1311.2936E} are clearly non-abelian. The value of the topological entanglement entropy is also within numerical error of $\log \sqrt{5} = 0.804719$ - the value for an abelian state at filling $2/5$, even though the state is non-abelian and the value of the total quantum dimension computed from the $S$-matrix leads to a clearly different value of $1.44768$ for the vacuum sector and a quantum dimension of the $\sigma$ particle of $(1+\sqrt{5})/2$. Note that a similar result was observed for classical stringnet models\cite{Hermanns-PhysRevB.89.205107}, where the constant correction to the entanglement only probes the abelian sector. Moreover, due to the small range of accessible perimeter values, we cannot probe any logarithmic correction to the area law which might emerge in some critical model\cite{Fradkin-PhysRevLett.97.050404,Oshikawa-2010arXiv1007.3739O}. While the resemblance of the entanglement entropy with that of an abelian $2/5$ state might not be numerically surprising - considering the extremely large $(>93\%)$ overlap of this state with the abelian Jain state at the same filling factor - the abelian quantum dimension of the purportedly non-abelian quasiholes is impossible to reconcile. This is strong evidence that the Gaffnian state is not gapped in the bulk, which casts a doubt on  the validity of our results for the Gaffnian entanglement entropy and quantum dimensions. Indeed our calculation involves making the cylinder infinitely long before extrapolating for large perimeters $L$. For a gapless state it is unclear whether this yields the same result as sending both length and perimeter to infinity while keeping a finite aspect ratio. Nevertheless this is an indication of the gapless nature of the Gaffnian, which we are going to confirm and quantify by looking for long range correlations. 

We now numerically compute the bulk correlation length in the Laughlin, Moore-Read and Gaffnian states. This correlation length is intrinsically related to the gap of the transfer matrix\cite{Fannes-1992}. The correlation function of an operator ${\cal O}(x)$ takes the form:
\beq
\langle {\cal O}^\dagger(x) {\cal O}(0) \rangle - \langle {\cal O}^\dagger(x)\rangle \langle {\cal O}(0) \rangle \propto e^{- |x|/\zeta}
\eneq where the correlation length $\zeta (L)= 2 \pi l_B^2 / (L\log(\frac{\lambda_1}{\lambda_2}))$ can be expressed in terms of the largest and second largest eigenvalues of the MPS transfer matrix  $\lambda_1(L)$ and $\lambda_2(L)$.  $L$ is the circumference of the cylinder where the correlation function is computed and $l_B$ is the magnetic length. In all the FQH states we considered we observed that the MPS transfer matrix was always gapped at finite $L$, which leads to a finite correlation length $\xi(L)$. When describing a gapped physical system, the correlation length $\zeta$ must remain finite in the thermodynamic limit $L\to \infty$. In Fig.~\ref{corrlengthlaughlinmr} we show the correlation lengths for the $\nu =1/3$ and $\nu = 1/5$ Laughlin states. As expected, they quickly saturate to a constant value, confirming the gapped nature of the Laughlin states. Fig.~\ref{corrlengthlaughlinmr} also provides the correlation length of the Moore-Read wavefunction. In this situation, the MPS transfer matrix is block diagonal in the abelian ($1, \psi$) and non-abelian ($\sigma$) sector  \cite{estienne-2013arXiv1311.2936E}. Hence two gaps exist, corresponding to the two correlation lengths of the abelian and non-abelian sectors. Both correlations lengths are finite in the thermodynamic limit and have roughly the same value as observed in Fig.~\ref{corrlengthlaughlinmr}. Ref.~\onlinecite{Baraban-PhysRevLett.103.076801} has found a correlation length of resp. $\zeta\simeq 2.7 l_B$ for the Moore-Read state in the vacuum sector, in agreement with our results. 

\begin{figure}[htb]
\includegraphics[width=0.9\linewidth]{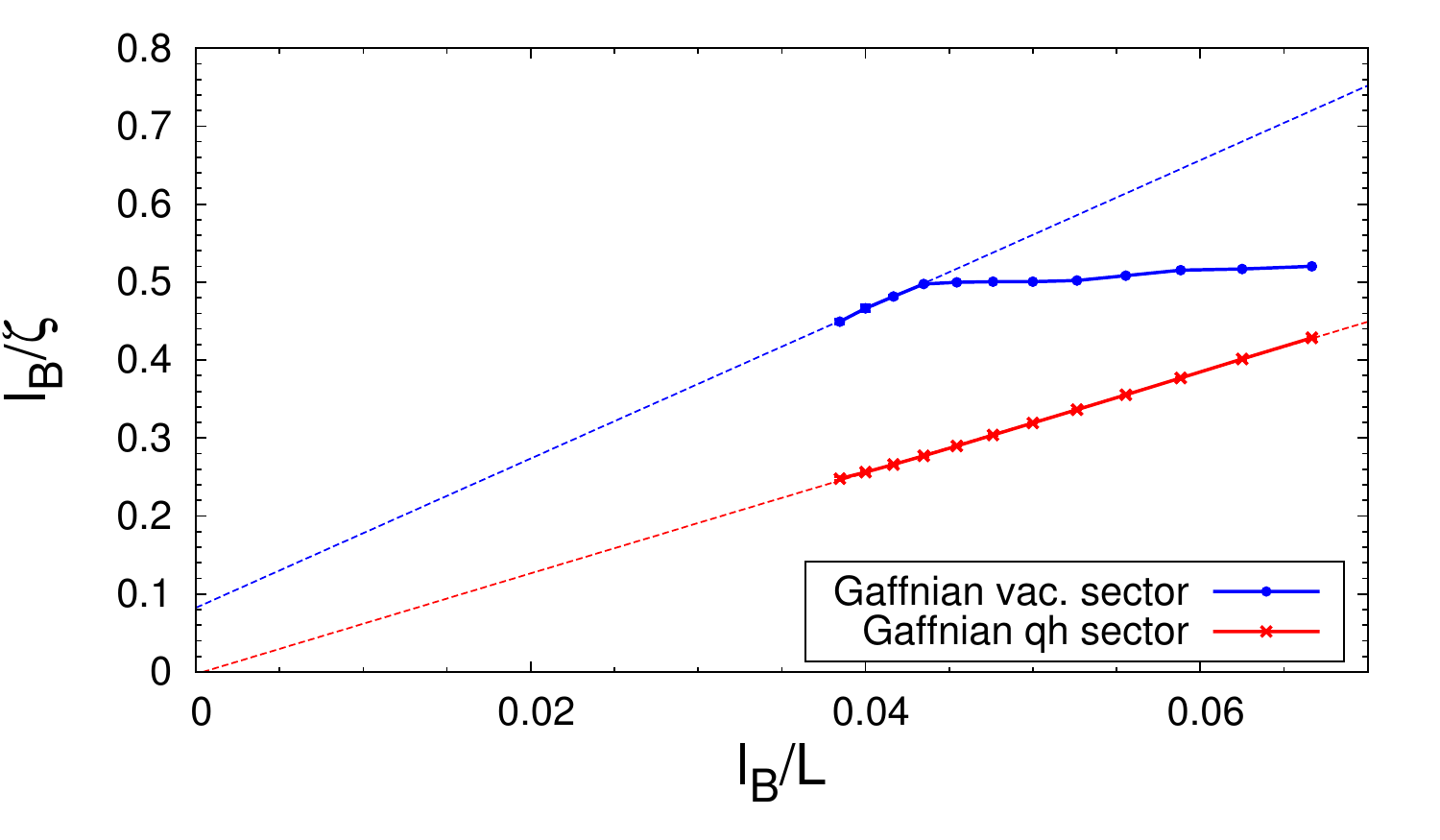}
\caption{Correlation lengths $\zeta$ of the Gaffnian state for both the vacuum and the quasihole sectors as a function of cylinder perimeter $L$. A linear fit as a function of $1/L$ for the correlation lengths $\zeta_{\rm qh}$ in the quasihole sector leads to an divergent extrapolated value. In the vacuum sector, due to the level crossing, it is difficult to make any reliable extrapolation.}
\label{corrlengthGaffnian}
\end{figure}

In Fig.~\ref{corrlengthGaffnian} we plot the correlation length for the non-unitary Gaffnian state for both sectors. For the quasihole (the field of scaling dimension $-1/20$ in the neutral CFT\cite{estienne-2013arXiv1311.2936E}) in the thermodynamic limit the correlation length diverges, signaling gaplessness. In the vacuum sector, the correlation length does not have a smooth behavior. While for small cylinder perimeter $L$, the correlation length seems to clearly extrapolate to a finite value, for larger $L$ the slope changes dramatically. This is due to a level crossing in the second transfer matrix eigenvalue $\lambda_2$. Note that the system size where the level crossing occurs is out of reach of previous techniques which are limited to sizes around $\simeq 17 l_B$. While other works using the exact expression of the Gaffnian quasihole states in terms of Jack polynomials\cite{Bernevig-2012arXiv1207.3305B} have indicated that the state fails to screen the non-abelian quasihole, this is the first clear calculation of the gapless nature of the non-unitary Gaffnian state.

In this paper we have computed the topological entanglement entropy for the Moore-Read, $\mathbb{Z}_3$ Read-Rezayi and Gaffnian states for both the vacuum and the quasihole sectors. While for unitary states the total quantum dimension and the quantum dimension of the individual quasiholes matches the CFT predictions, for the Gaffnian state, we find quantum dimensions identical to those of the abelian FQH state at identical filling. We also computed the correlation lengths in these states and find that the unitary states have finite correlation lengths in the thermodynamic limit, while the non-unitary Gaffnian has diverging correlation length in at least the quasihole sector, signaling gaplessness.

\emph{Acknowledgements}
We thank M. Zaletel, J. Dubail, P. Bonderson, T. Grover, E. Ardonne, F. Pollmann, J-B. Zuber, Z. Papic and Y.-L. Wu for discussions. BAB and NR were supported by NSF CAREER DMR-0952428, ONR-N00014-11-1-0635, MURI-130- 6082, NSF-MRSEC DMR-0819860, DARPA - N66001-11-1-4110, Packard Foundation, and Keck grant. NR was supported by the Princeton Global Scholarship.

\newpage

\bibliography{../mpsnonunitarity.bib}

\begin{center}
{\bf Supplementary Material}
\end{center}

\begin{figure}[htb]
\includegraphics[width=0.9\linewidth]{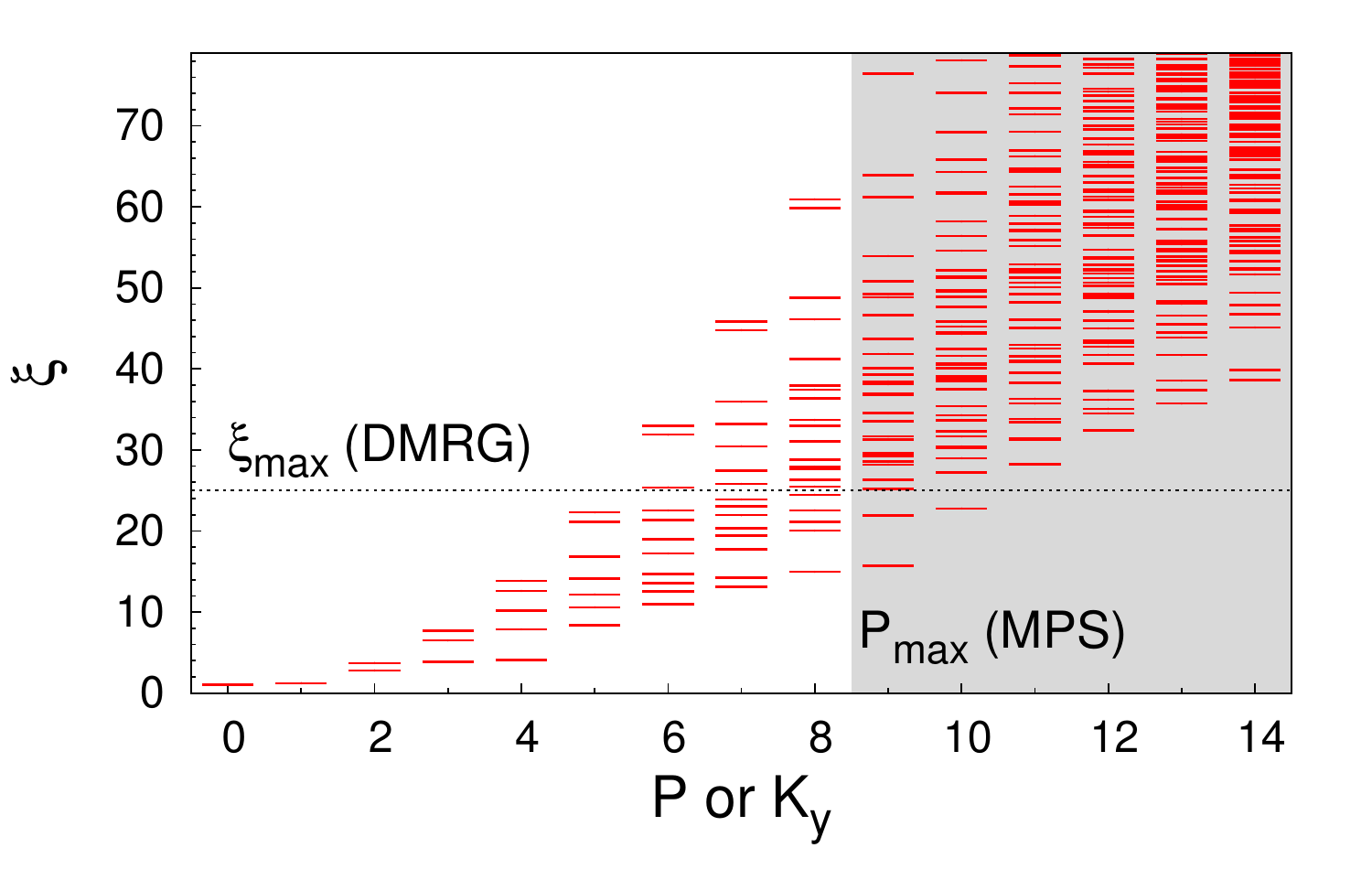}
\caption{Schematic description of the difference between the truncation of the DMRG algorithm and the one of the exact MPS. Here we show the entanglement spectrum of the Laughlin state $\nu=1/3$, with the entanglement energies $\xi$ as a function of momentum $K_y$ (equivalent to the CFT level $P$) along the cylinder perimeter. The DMRG algorithm approximates by conserving levels below the entanglement energy cut-off $\xi_{\rm max}$ (depicted here by an horizontal dotted line). The MPS truncation keeps all the levels of the entanglement spectrum up to the momentum $P_{\rm max}$ and discards the levels beyond this point (here the gray area).}
\label{DMRGvsMPS}
\end{figure}

In this Supplementary Material, we provide additional numerical results that might be relevant to a more specialized audience. We first explain the truncation parameter in our numerical calculations. The meaning of the $P_{\rm max}$ truncation parameter can be understood by plotting the entanglement spectrum versus the total momentum of the particles in the $A$ region as in Fig.~\ref{DMRGvsMPS}. Our approximation takes into account all the levels in the entanglement spectrum from zero up to a $P_{\rm \max}$ momentum. We contrast this with the DMRG approximation, where only levels up to a certain "entanglement energy"\cite{li2008} are kept. For a chiral state spectrum, as ours, the difference between the two approximations is expected to be minimal. Indeed, one expects that the entanglement spectrum mimics the energy spectrum of the system chiral modes leading roughly to a linear relation between the momentum truncation and the ``entanglement energy'' truncation.

\begin{figure}[htb]
\includegraphics[width=0.9\linewidth]{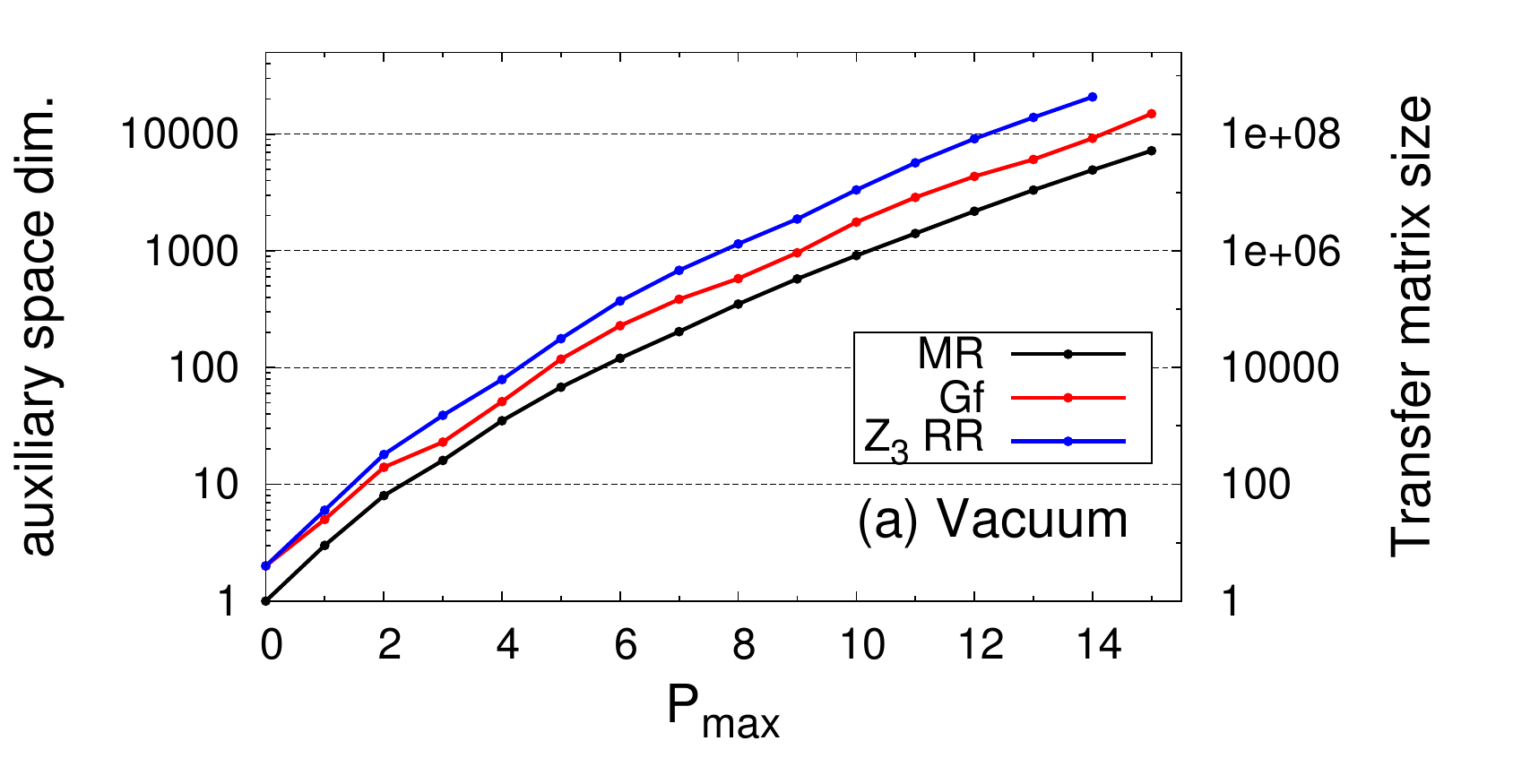}
\includegraphics[width=0.9\linewidth]{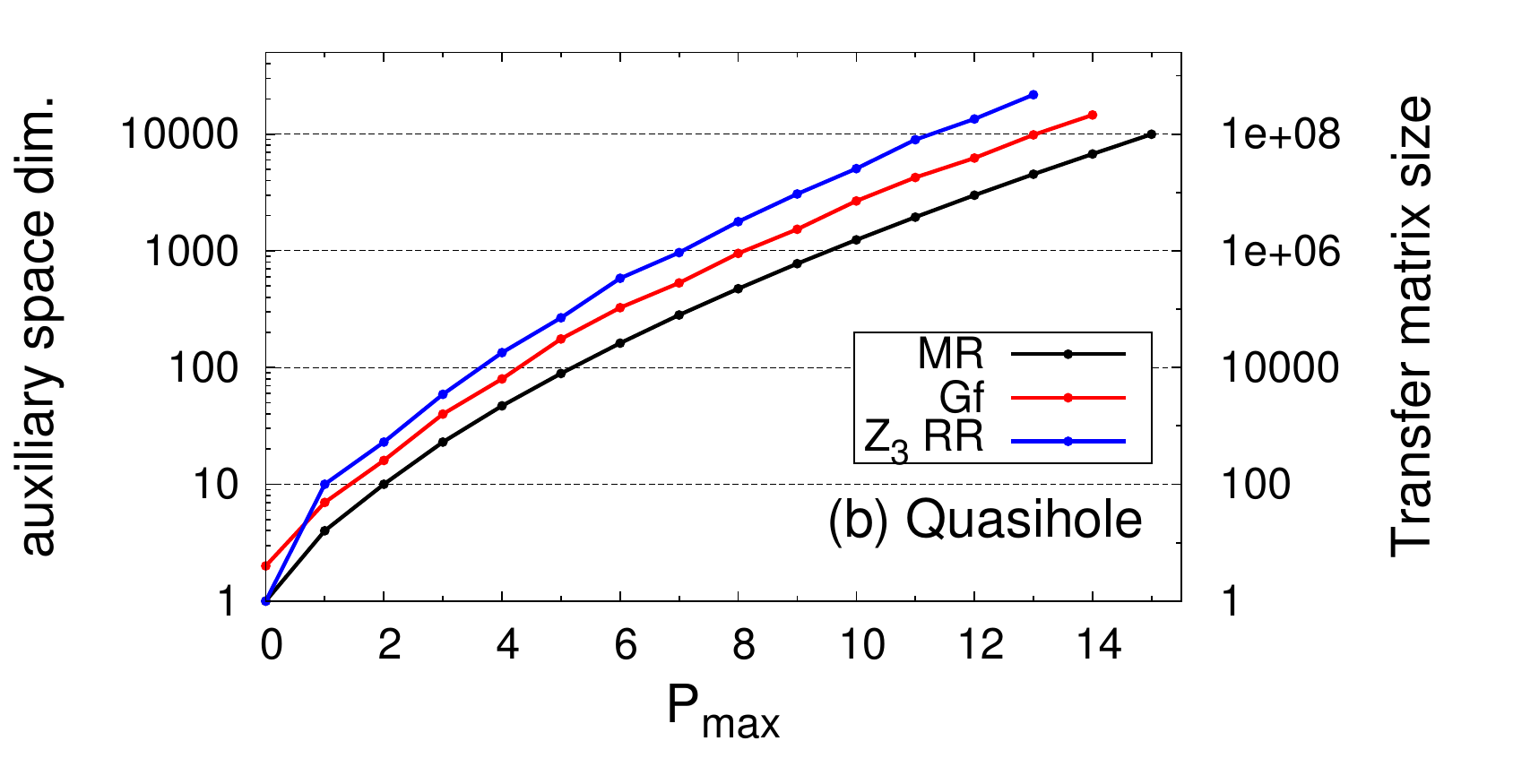}
\caption{Dimensions of the auxiliary space for the Moore-Read, Gaffnian and $\mathbb{Z}_3$ Read-Reazayi states as a function of the truncation parameter $P_{\rm max}$ in (a) the vacuum sector (top panel) and (b) the quasihole sector (lower panel). The left vertical axis gives the auxiliary space dimension, while the right vertical axis shows the size of the transfer matrix (\emph{i.e.} the square of the auxiliary space dimension).}
\label{auxilarydim}
\end{figure}

\begin{figure}[htb]
\includegraphics[width=0.9\linewidth]{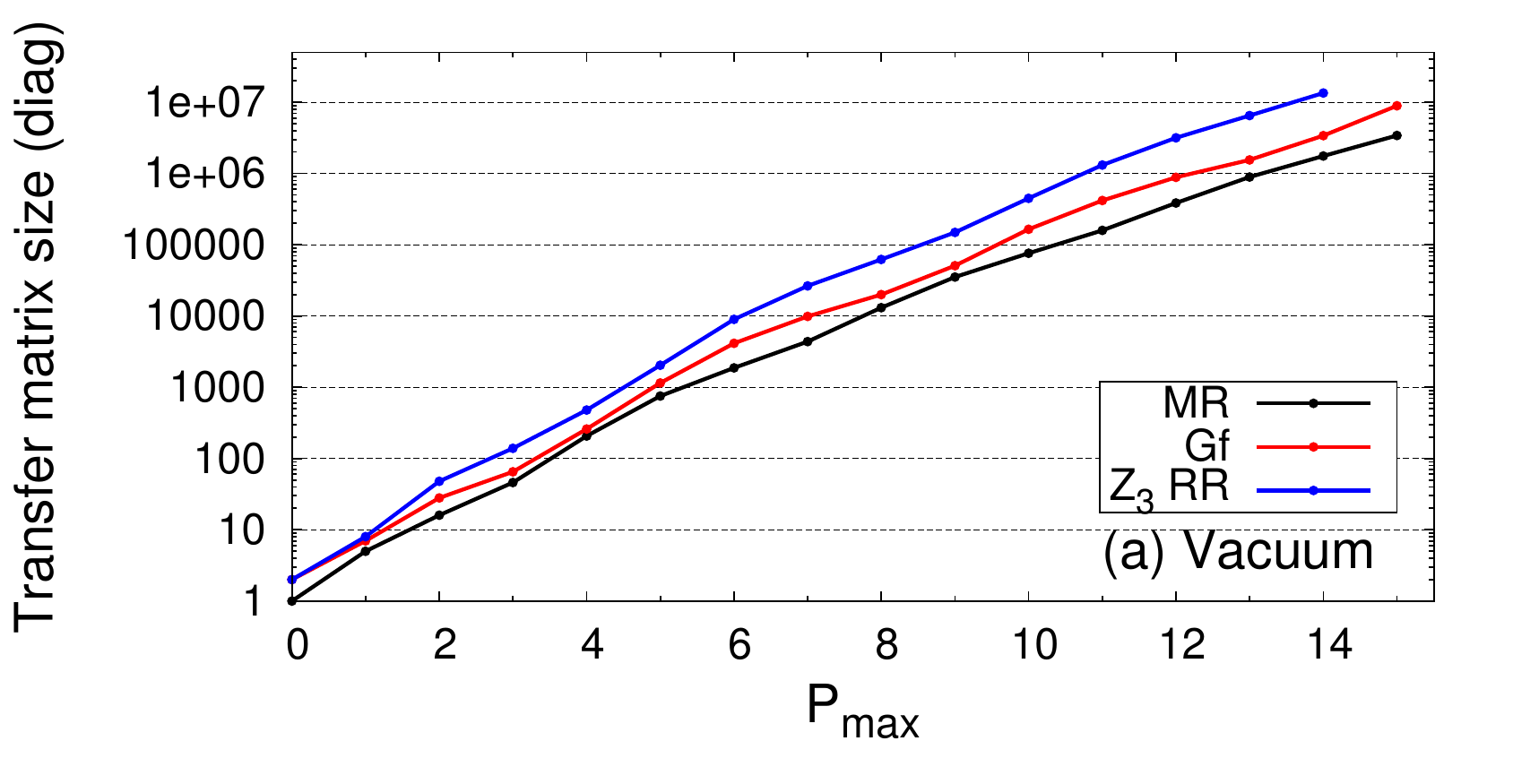}
\includegraphics[width=0.9\linewidth]{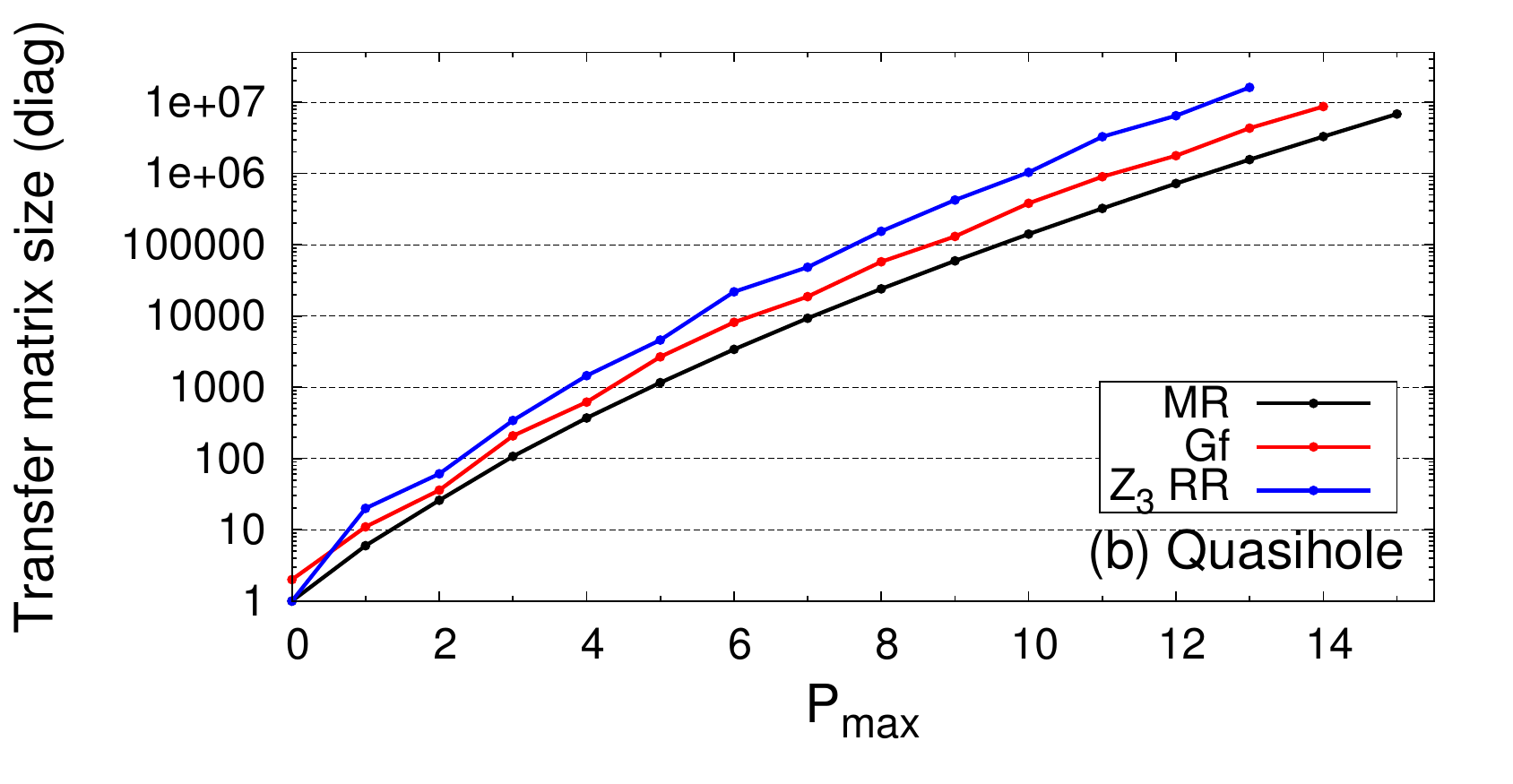}
\caption{Effective dimension of the transfer matrix when computing the largest eigenstate for the Moore-Read, Gaffnian and $\mathbb{Z}_3$ Read-Reazayi states as a function of the truncation parameter $P_{\rm max}$ in (a) the vacuum sector (top panel) and (b) the quasihole sector (lower panel).}
\label{auxilarydimdiag}
\end{figure}

The size of the auxiliary space depends on the state, $P_{\rm max}$, and the topological sector. In Ref.~\onlinecite{estienne-2013arXiv1311.2936E}, we have discussed the size of the auxiliary space for the Laughlin, Moore-Read and Read-Rezayi state in the vacuum sector (see Fig.~4 in Ref.~\onlinecite{estienne-2013arXiv1311.2936E}). For sake of completeness, we provide the dimension of the auxiliary space for the Gaffnian state in the vacuum sector and the all the states that we have considered in this letter in the quasihole sector. The dimensions for the auxiliary space and the size of the transfer matrix are shown in Fig.~\ref{auxilarydim}. These values take into account the reduction from the trimming procedure and the restriction to a single topological sector. When computing the left and right eigenstates associated to the largest eigenvalue, one can use the diagonal part of the transfer matrix that preserves the $U(1)$ charge and conformal dimension as discussed in Ref.~\onlinecite{estienne-2013arXiv1311.2936E}. This leads to reduced effective transfer matrices whose corresponding sizes are given in Fig.~\ref{auxilarydimdiag}.

The topological entanglement entropy is related to the quantum dimensions $d_a$  of each topological sector. In Table~\ref{table:quantumdimension}, we summarize the values derived from the CFT for each of the model states we have considered. The Verlinde formula\cite{Verlinde1988360} can be used to extract the quantum dimensions from the modular $S$-matrix
\begin{equation}
d_a= \frac{S_{a,o}}{S_{0,o}}
\end{equation}
where $a,b$ are the field indices, $0$ stands for the identity field, and $o$ is the unique field such that $S_{a,o} > 0$ for all $a$ (\emph{i.e.} $o$ is the Perron-Froebenius vector of the fusion matrices). It can be shown that $o$ is the field with the lowest conformal dimension\cite{Gannon2003335}. For a unitary CFT we recover the standard result $d_a= S_{a,0}/S_{0,0}$. 

For the MR state, they can be derived directly form the $S$-matrix of the neutral part \emph{i.e.} the Ising model. The derivation of these values for the generic $\mathbb{Z}_k$ RR states are more tedious and can be found in Ref.~\onlinecite{Fendley-JStatPhys2007}. For the Gaffnian state, one should pay attention that the column of the $S_{0,o}$ entries are not related to the identity but to one of the quasihole fields (namely $o=\phi$).

\begin{table}[htbp]
\centering
\begin{tabular}{ l |c|c|c }
Model state & $d_{\rm vac}$ & $d_{\rm qh}$ & $\cal D$\\
\hline
\hline
MR & $1$ & $\sqrt{2}$ & $2\sqrt{2}$ \\
\hline
$\mathbb{Z}_3$ RR &  $1$ & $\frac{1+\sqrt{5}}{2} $ & $\frac{5}{2 \sin \left(\frac{\pi}{5}\right)}$ \\
\hline
Gaffnian & $1$ & $\frac{1+\sqrt{5}}{2}$ & $\frac{5}{2 \sin \left(\frac{\pi}{5}\right)}$
\end{tabular}
\caption{Quantum dimensions for the MR, $\mathbb{Z}_3$ RR and Gaffnian states. $\cal D$ is the total quantum dimension.}\label{table:quantumdimension}
\end{table}

\begin{figure}[htb]
\includegraphics[width=0.9\linewidth]{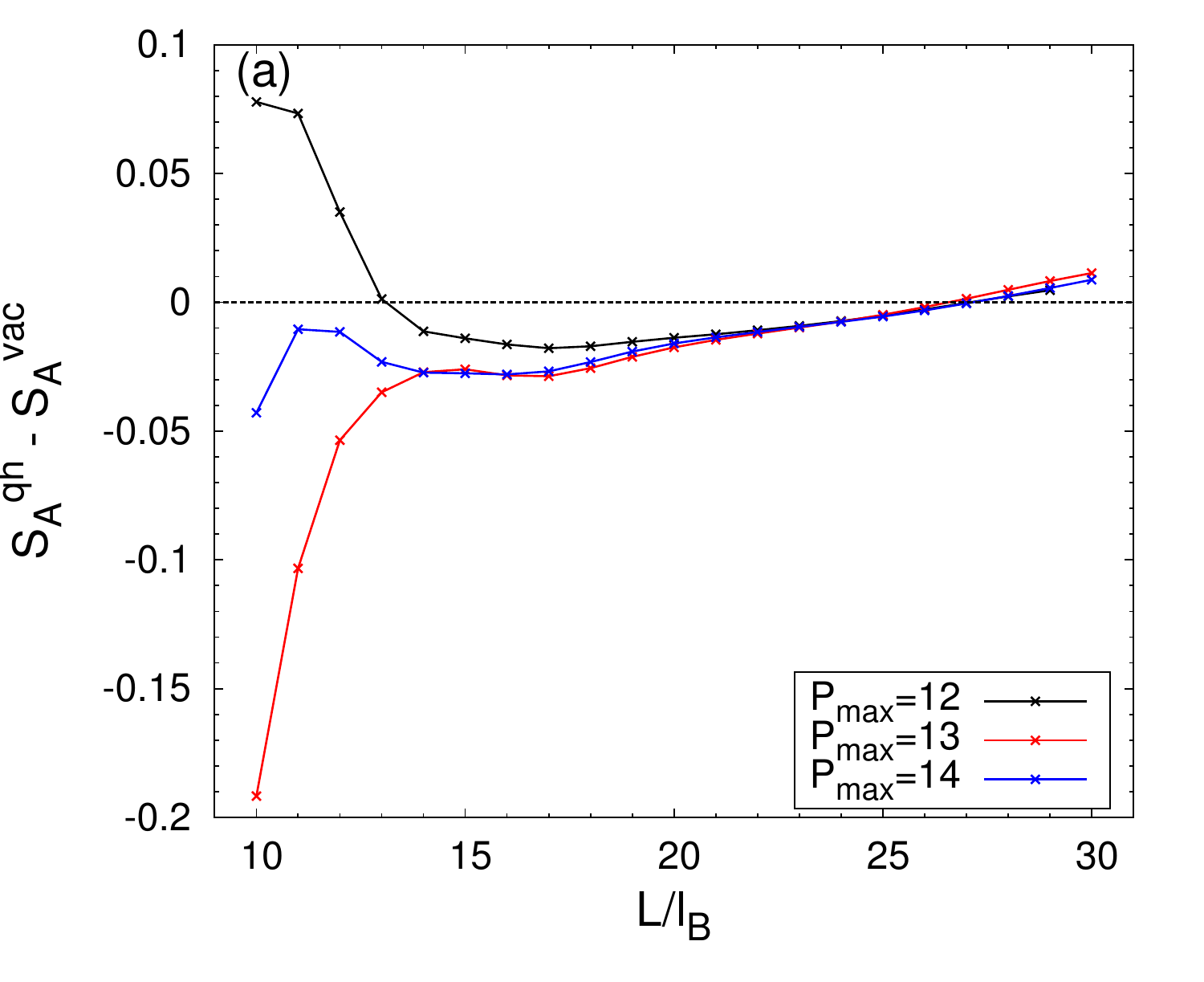}
\includegraphics[width=0.9\linewidth]{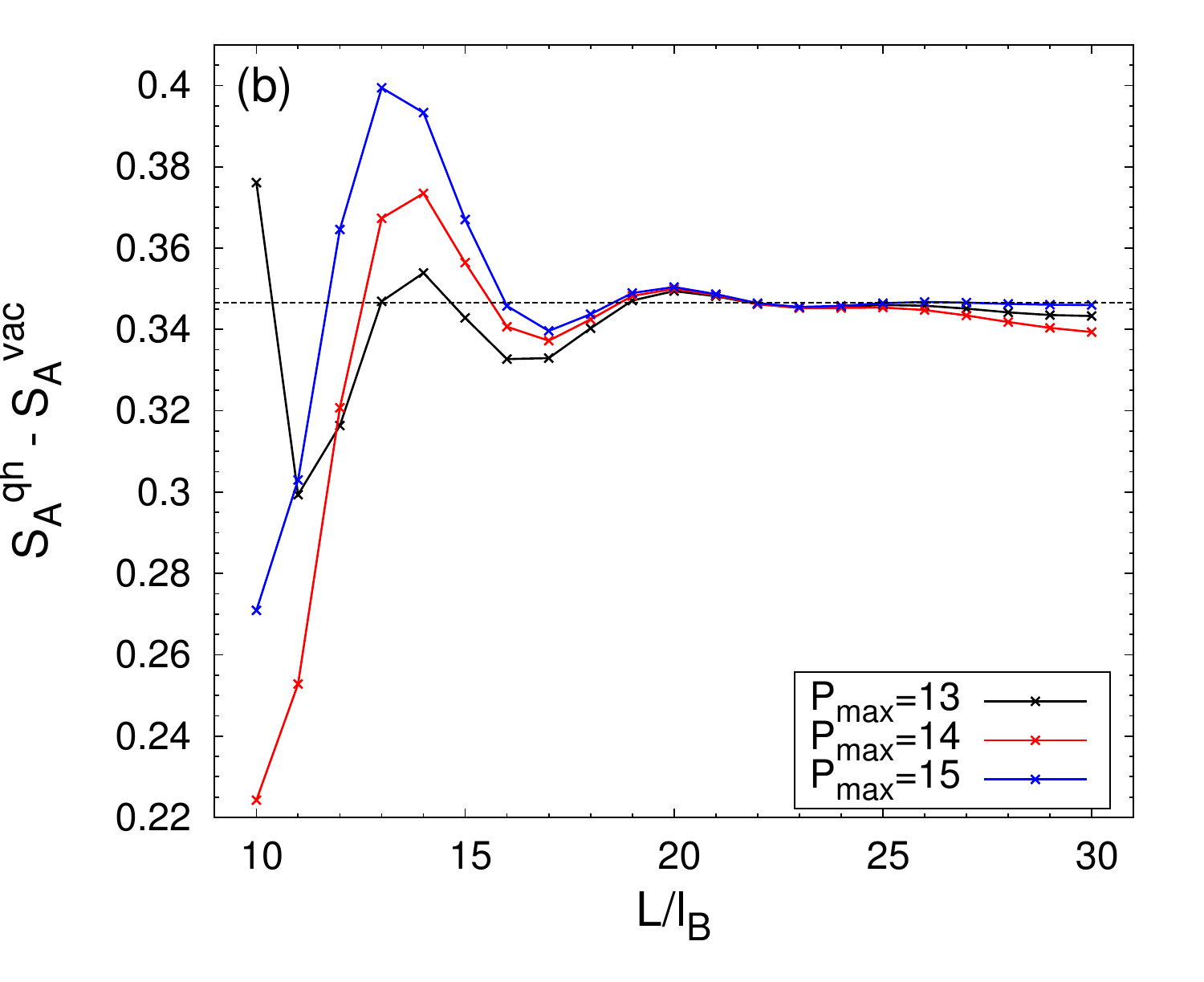}
\caption{Difference of the entanglement entropies between the quasihole sector ($S_A^{\rm qh}$) and the vacuum sector ($S_A^{\rm vac}$) as a function of the cylinder perimeter $L$  and the three largest truncation parameters $P_{\rm max}$ that can be reached. {\it Upper panel} : difference for the Gaffnian state. The dashed line is a guide for the eye of what should be this difference if the quantum dimension for the quasihole was one. {\it Lower panel} : difference for the Moore-Read state. The dashed line is a guide for the eye of what should be this difference in the thermodynamic limit, i.e. $\ln(2)/2$.}
\label{diffentropy}
\end{figure}

In the article, we have computed the entanglement entropy in both the vacuum and quasihole sectors. As expected and as observed, the area law linear factor of the entanglement entropy is identical for both sectors in the three cases we have considered (namely the Moore-Read state, the Gaffnian state and the  $\mathbb{Z}_3$ Read-Rezayi state). This implies that the difference in the entanglement entropy between the quasihole and the vacuum sectors should be dominated by a constant term, the difference of the topological entanglement entropies $\ln(d_{\rm qh})$. In Fig.~\ref{diffentropy}, we plot this difference for the Gaffnian state (Fig.~\ref{diffentropy}a) and the Moore-Read state (Fig.~\ref{diffentropy}b). For the Moore-Read state, the difference converged towards $\ln(d_{\rm qh})=\ln\left(\sqrt{2}\right)$. For the Gaffnian state, the difference is clearly away from $\ln\left(\frac{1+\sqrt{5}}{2}\right)\simeq 0.48$, and more compatible with a quantum dimension for the quasihole being equal to one.

\begin{figure}[htb]
\includegraphics[width=0.9\linewidth]{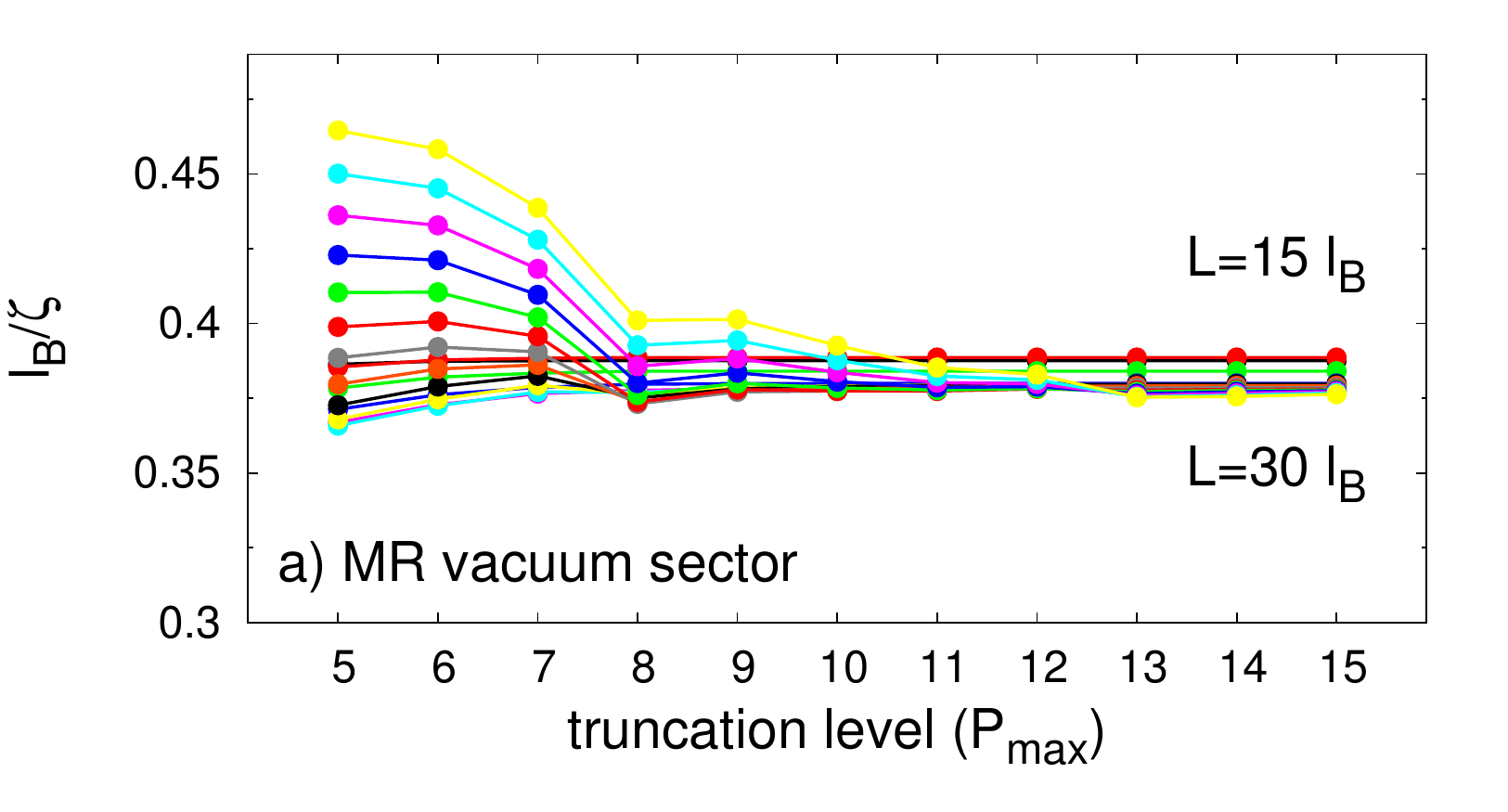}
\includegraphics[width=0.9\linewidth]{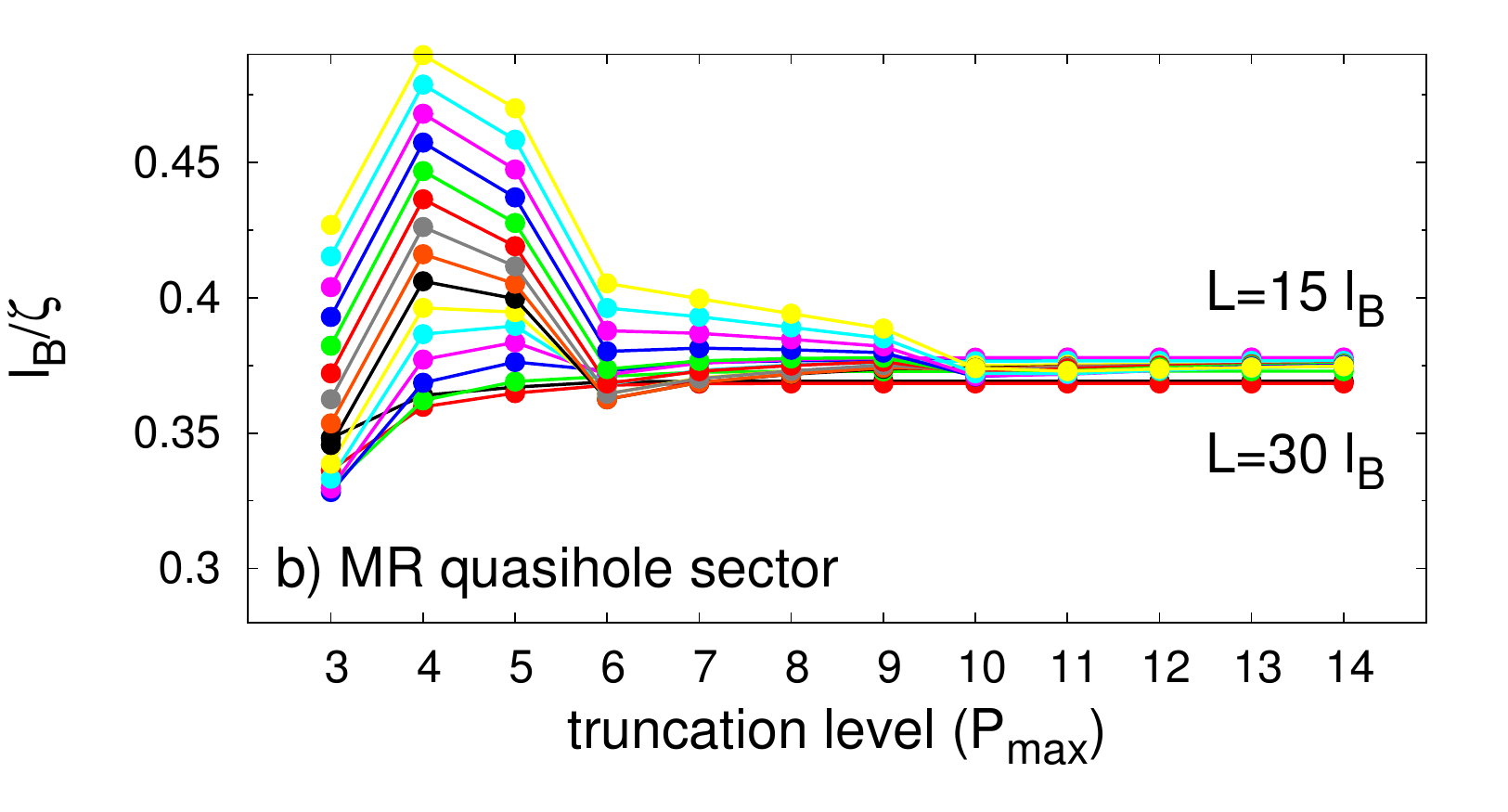}
\caption{Correlation length $\zeta$ for the Moore-Read state in the vacuum sector (top panel) and the quasihole sector (lower panel). The correlation length is computed for all truncation levels between $P_{\rm max}=3$ to $P_{\rm max}=14$ ($P_{\rm max}=15$ for the vacuum sector) and for all perimeters from $L=15 l_B$ to $L=30 l_B$ by steps of $1 l_B$.}
\label{corrlengthmrperimeter}
\end{figure}

\begin{figure}[htb]
\includegraphics[width=0.9\linewidth]{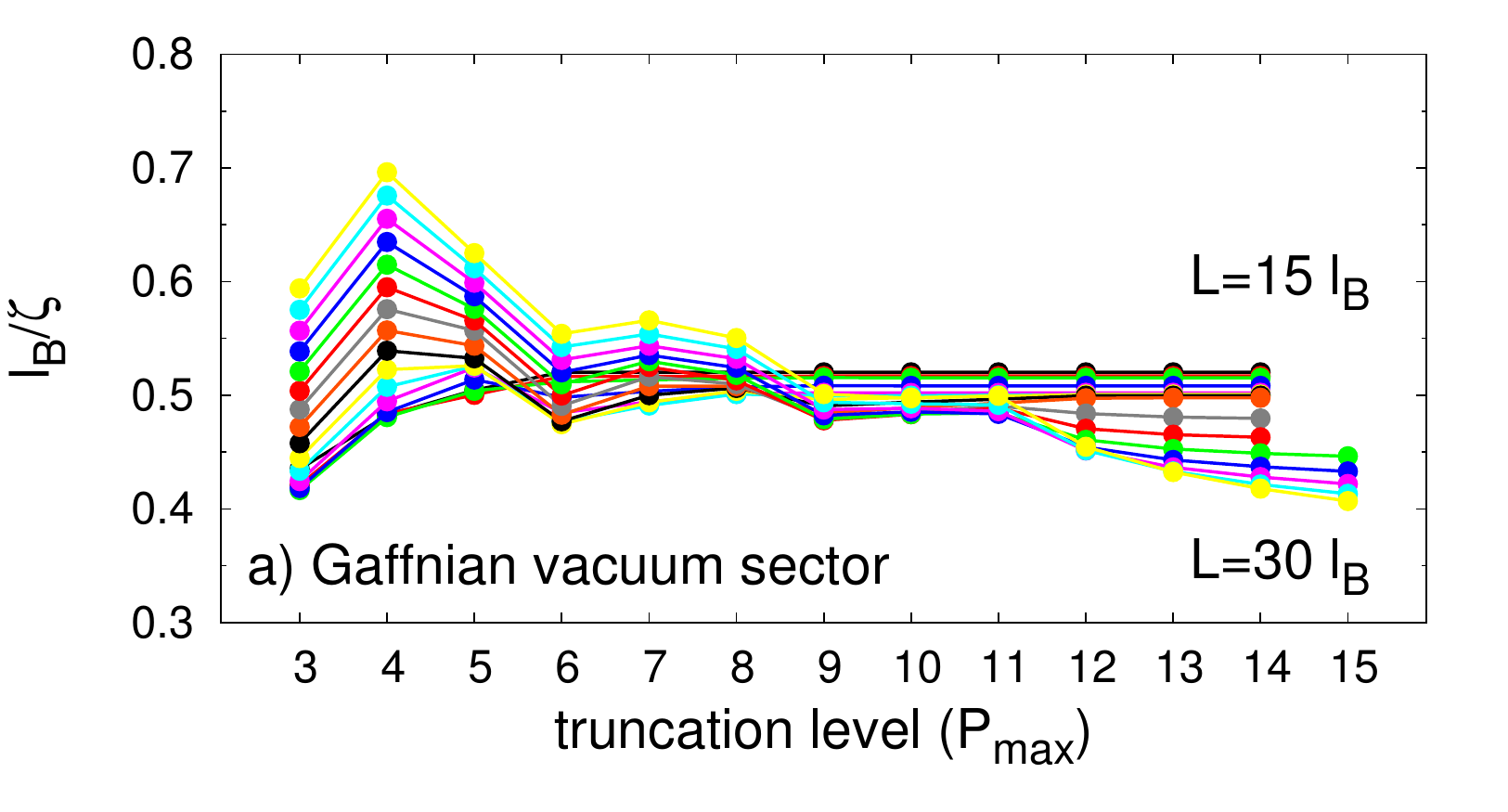}
\includegraphics[width=0.9\linewidth]{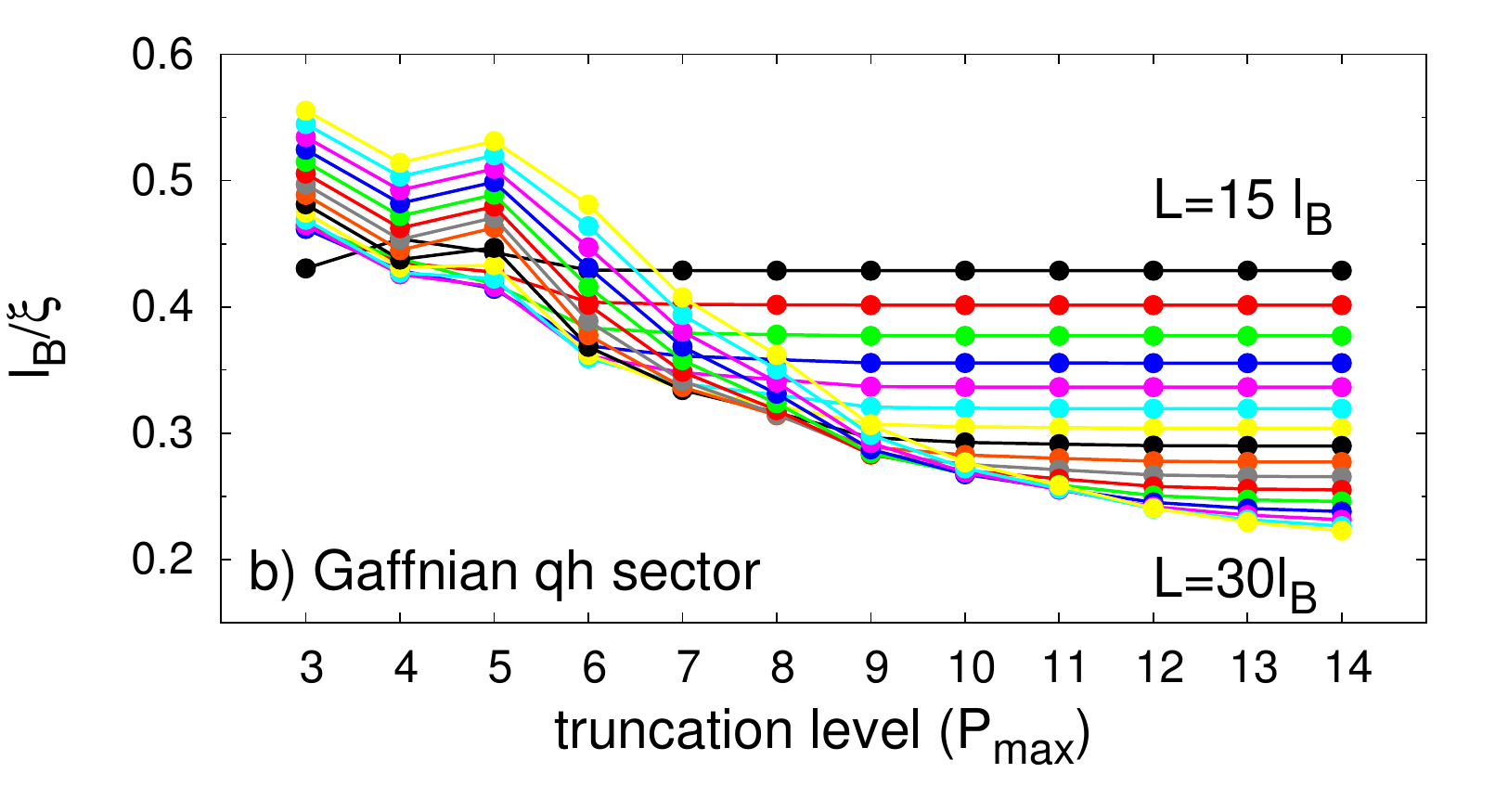}
\caption{Correlation length $\zeta$ for the Gaffnian state in the vacuum sector (top panel) and the quasihole sector (lower panel). The correlation length is computed for all truncation levels between $P_{\rm max}=3$ to $P_{\rm max}=14$ and for all perimeters from $L=15 l_B$ to $L=30 l_B$ by steps of $1 l_B$. Note that for the vacuum sector and $P_{\rm max}=15$ , we have computed the correlation length only for $L \geq 26 l_B$.}
\label{corrlengthGaffnianperimeter}
\end{figure}

We now provide several plots showing the dependence of the correlation length, defined in the text, with respect to both the truncation parameter $P_{\rm max}$ and the circumference of the cylinder $L$. Here we focus on the Moore-Read state (Fig.~\ref{corrlengthmrperimeter}) and the Gaffnian state (Fig.~\ref{corrlengthGaffnianperimeter}) in both the vacuum and quasihole sectors. In the letter, we give the correlation lengths $\zeta$  as a function of $1/L$. We have only considered values of $L$ for which   the convergence  of $\zeta^{-1}$ as a function of $P_{\rm max}$ was better than $10^{-2}$. In the absence of a reliable extrapolation as a function of $P_{\rm max}$, the value of $\zeta^{-1}$ averaged over the three largest $P_{\rm max}$ values. As can be observed in Figs.~\ref{corrlengthmrperimeter} and~\ref{corrlengthGaffnianperimeter}, such an approximation might slightly overestimates $\zeta^{-1}$ if the convergence is slow.

\begin{figure}[htb]
\includegraphics[width=0.9\linewidth]{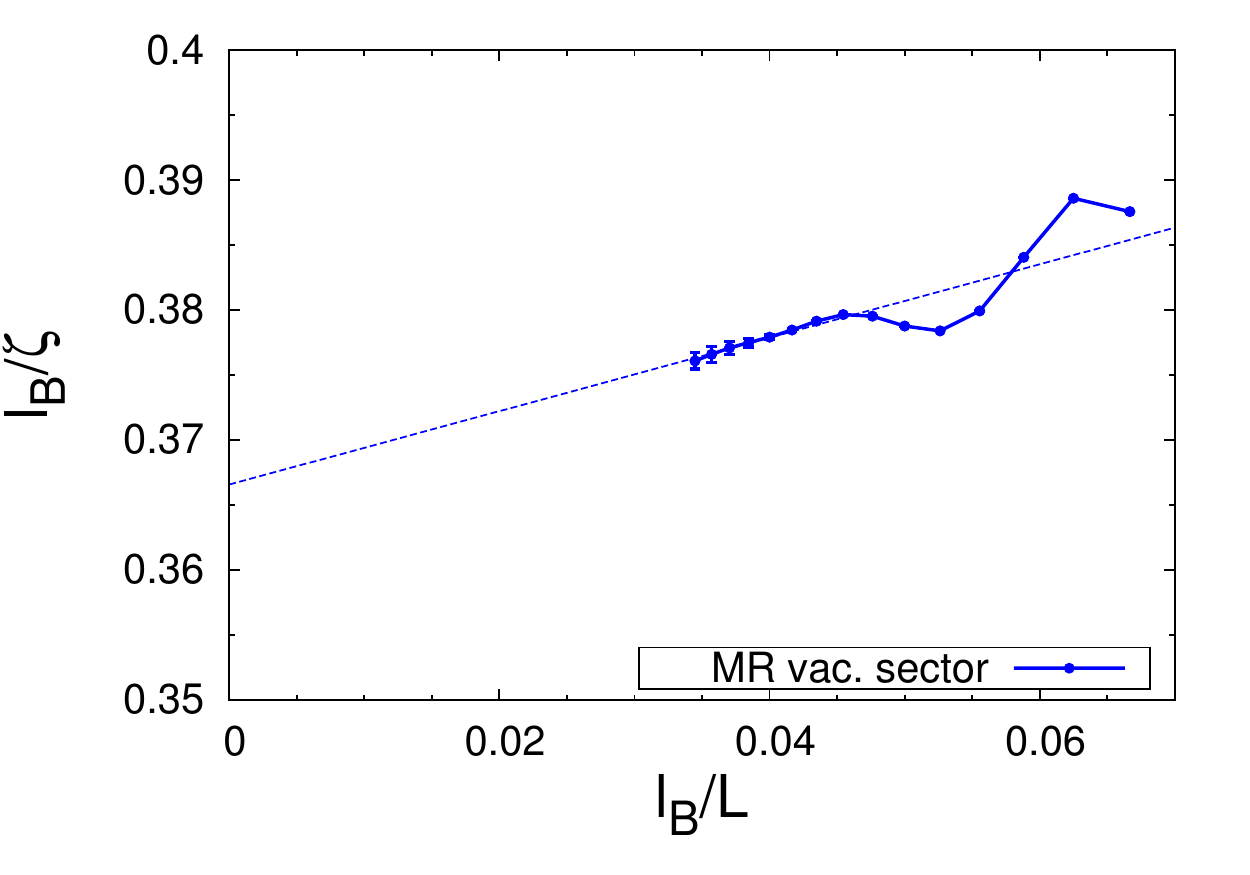}
\caption{Correlation lengths $\zeta$ of the Moore-Read state for an off-diagonal block $E_{a,b}$, where $a$ is the vacuum sector and $b$ is the $(e/2)$ abelian quasihole sector. A linear fit as a function of $1/L$ gives the thermodynamical values $\zeta_{\rm vac}/l_B=2.73(1)$ in agreement with the correlation length extracted in the diagonal $U(1)$ sector.}
\label{corrlengthmrperimeteroffdiag}
\end{figure}

Finally, we would like to address the question of the correlation length related to a change of topological sector. The $B$ matrices are block diagonal with respect to the topological sector. We denote $B^{m}_a$ the block of the $B^{m}$ matrix (where $m=0,1$ is the physical index corresponding to an empty or occupied orbital) corresponding to the topological sector $a$. The transfer matrix $E$ is defined as 
\begin{eqnarray} 
E=\sum_{a,b} E_{a,b} &\;\; {\rm with} \;\;& E_{a,b}=\sum_{m} B^{m}_a \otimes \left(B^{m}_b\right)^{*}
\end{eqnarray}
The calculations of the entanglement entropy in a given topological sector $a$ that we have discussed in this letter involves only the largest left and right eigenstates of the diagonal block $E_{a,a}$ of the transfer matrix. On the other hand if we want to claim that the correlation length is finite for any type of operator (including those that induces a topological sector change), we need the full transfer matrix to be gapped. In Fig.~\ref{corrlengthmrperimeteroffdiag}, we extract the correlation length from an off-diagonal block of the MR state transfer matrix $E_{a,b}$, where the left topological sector $a$ is the vacuum, while the right topological sector $b$ is the $e/2$ abelian quasihole. As expected, the value of $\zeta$ is in agreement with those extracted from the diagonal blocks of $E$. The other off-diagonal blocks, which relates sectors that differ by a non-Abelian quasihole, cannot be diagonalized and thus are not relevant in this discussion.

\end{document}